%% file: eprint.tex
\documentclass[12pt]{article}
\usepackage{graphicx}
\usepackage{lineno}
\usepackage{url}

\newcommand\pubnumber{ATLAS-CMS}
\newcommand\pubdate{\today}

\def\institute{Deutsches Elektronen-Synchrotron}
\def\authemail{\footnote{Contact: hugo.alberto.becerril.gonzalez@cern.ch}}

\def\Title#1{\begin{center} {\Large #1 } \end{center}}
\def\Author#1{\begin{center}{ \sc #1} \end{center}}
\def\Address#1{\begin{center}{ \it #1} \end{center}}

\newcommand\pubblock{\rightline{\begin{tabular}{l} \pubnumber\\
         \pubdate  \end{tabular}}}
\newenvironment{Abstract}{\begin{quotation}  }{\end{quotation}}
\newenvironment{Presented}{\begin{quotation} \begin{center} 
             PRESENTED AT\end{center}\bigskip 
      \begin{center}\begin{large}}{\end{large}\end{center} \end{quotation}}

\input econfmacros.tex

\begin{document}
\begin{titlepage}
\pubblock
\footnote{Copyright 2023 CERN for the benefit of the ATLAS and CMS Collaborations. Reproduction of this article or parts of it is allowed as specified in the CC-BY-4.0 license}
\vfill
\Title{Top properties (excluding mass) and ancillary measurements}
\vfill
\Author{Hugo Alberto Becerril Gonzalez\authemail \\ on behalf of the ATLAS and CMS collaboration}
\Address{\institute}
\vfill
\begin{Abstract}
A review on recent top quark properties measurements by the ATLAS and CMS Collaborations in pp collisions at the LHC is presented.
\end{Abstract}
\vfill
\begin{Presented}
$16^\mathrm{th}$ International Workshop on Top Quark Physics\\
(Top2023), 24--29 September, 2023
\end{Presented}
\vfill
\end{titlepage}
\def\thefootnote{\fnsymbol{footnote}}
\setcounter{footnote}{0}

\section{Introduction}

The top quark, characterized by its large mass of approximately 172.5 GeV, occupies a unique position within the Standard Model (SM) and exhibits relevance in various areas of physics, as indicated by numerous Beyond the Standard Model theories (BSM). Due to its exceptional mass, the top quark experiences decay prior to the process of hadronization, which allows for a direct exploration of its properties through the analysis of its decay products. Leveraging the extensive LHC Run-2 dataset, of roughly 135 fb$^{-1}$ of proton-proton collisions at a center-of-mass energy of 13 TeV (per experiment), presents an outstanding opportunity for conducting meticulous investigations into the attributes of the top quark. In the forthcoming discussion, we will mention recent findings from both the ATLAS \cite{The_ATLAS_Collaboration_2008} and CMS \cite{The_CMS_Collaboration_2008} experiments.

Top quarks usually decay via $t \rightarrow W b$, so the $t\bar{t}$ final states is determined by the $W$ boson decay that can be either hadronically ($W \rightarrow q\bar{q}$) or leptonically ($W \rightarrow l\nu$) with $l = e$ or $\mu$. The measurements discussed here were performed with at least one leptonically decaying $W$ boson in its final state. 

To record potentially $t\bar{t}$ events with the detectors, special single lepton triggers with high transverse momentum ($p_{T}$) are used. The electrons are reconstructed using a combination of the calorimeters and the tracker system while muons are mostly reconstructed using the information from the tracking systems.  

Hadrons do not leave a clear signature in the detectors; these are reconstructed as jets using the anti-kT clustering algorithm \cite{Matteo} with a distance parameter $R=0.4$. Sometimes, larger jets ($R = 0.8$ or $R = 1.0$) are needed to reconstruct boosted hadronically decaying top quarks along with dedicated top-tagging algorithms. The presence of one (or two) neutrinos in the final state is observed as missing transverse momentum. To perform the reconstruction of the four momentum of the top quarks, different techniques are employed utilizing available measurements of the decay products and constrains on the mass of the top quark and the $W$ boson.

\section{Evidence for the charge asymmetry in $pp \rightarrow t\bar{t}$ production at $\sqrt{s}$ = 13 TeV with the ATLAS detector}

The measurement of the top quark charge asymmetry is presented in \cite{Aad2023} with 139 fb$^{-1}$ proton-proton collisions using $t\bar{t}$ dileptonic and semi-leptonic events. SM predicts a subtle difference in the angular distribution of top quarks and antiquarks in $q\bar{q} \rightarrow t\bar{t}$, that could be enhanced by unknown physics \cite{PhysRevD.98.014003}. 

The measurement makes use of reconstruction techniques adapted to both the resolved and boosted topologies. To account for detector resolution and acceptance effects, a Bayesian unfolding procedure is performed. The combined inclusive $t\bar{t}$ charge asymmetry is determined to be $A_{C}$ = 0.0068 $\pm$ 0.0015, which deviates from zero by 4.7$\sigma$. 

\begin{figure}[!h!tbp]
\centering
\includegraphics[height=2.0in]{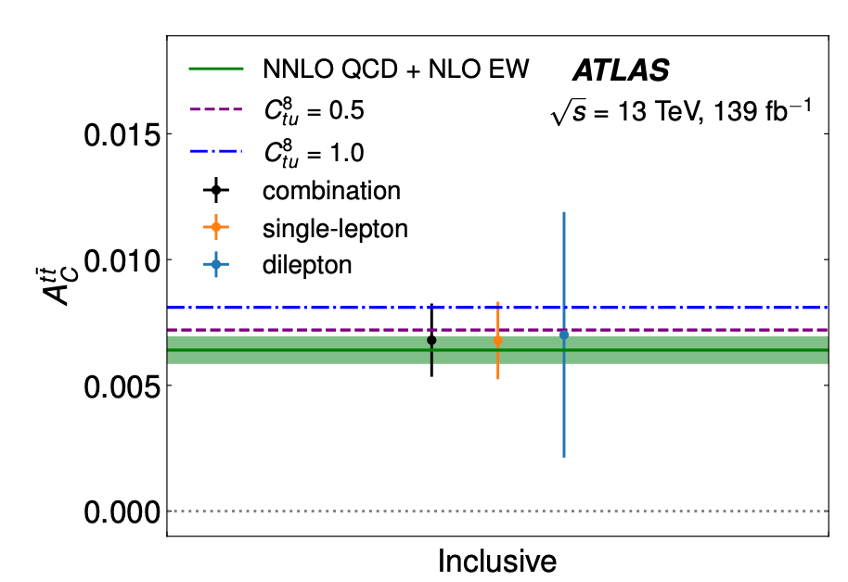}
\includegraphics[height=2.0in]{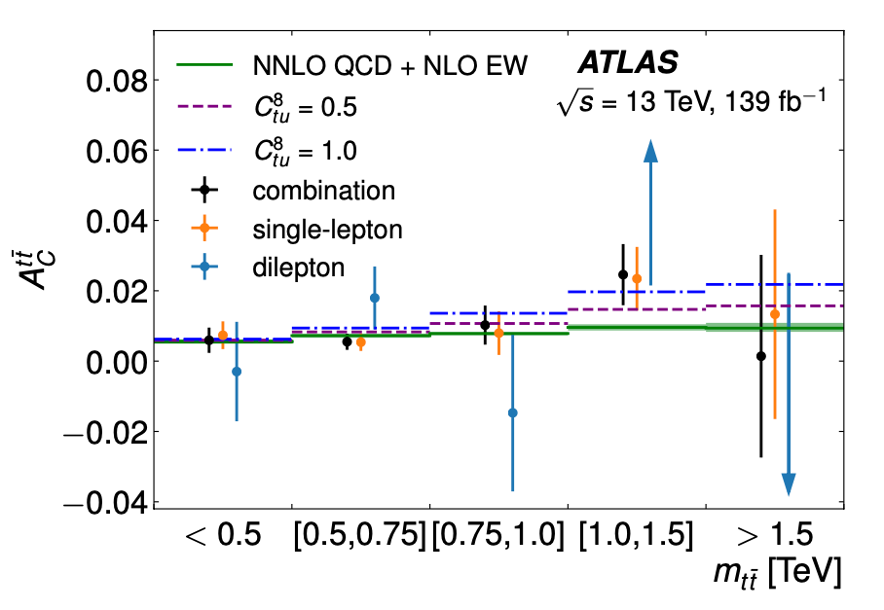}
\caption{The unfolded inclusive (left) and differential charge asymmetries as a function of the
invariant mass (right). SM theory predictions calculated at NNLO in QCD and NLO
in EW theory. The impact of the linear term of the $C^8_{tu}$ Wilson coefficient on the $A_C$ is shown as dashed lines \cite{Aad2023}.}
\label{fig:1}
\end{figure}

Differential measurements are performed as a function of the invariant mass, transverse momentum, and longitudinal boost of the $t\bar{t}$ system. Both the inclusive and differential measurements are found to be compatible with the SM predictions, at next-to-next-to-leading order in quantum chromodynamics perturbation theory with next-to-leading-order electroweak corrections (Fig. \ref{fig:1}). 

The measurements are also interpreted in the framework of the SMEFT, very valuable in disentangling blind directions in global SMEFT fits. Individual limits on the Wilson coefficients are derived from the inclusive charge asymmetry measurement and from the differential measurement versus $m_{t\bar{t}}$ making considerably improvement on the limits derived from the LHC 8 TeV \cite{Aaboud2018} combination and from the Tevatron combination \cite{PhysRevLett.120.042001}.

\section{Measurement of the charge asymmetry in top-quark pair production in association with a photon with the ATLAS experiment}

The $t\bar{t}$ charge asymmetry is affected by dilution at the LHC due to the substantial occurrence of gluon-gluon-initiated $t\bar{t}$ events (constituting nearly 90\% of the total) where there is no charge asymmetry at any order of perturbation. Nevertheless, it can be enhanced in other topologies, such as $t\bar{t} + \gamma$. The overall asymmetry in $t\bar{t} + \gamma$ events at a center-of-mass energy of $\sqrt{s}$ = 13 TeV is expected to exhibit a negative value, typically falling within the range of 1\% to 2\%, based on the specific phase space conditions, and additional factors stemming from BSM physics can potentially influence the results.

The measurement of the charge asymmetry in \cite{2023137848} is conducted in the single lepton channel. As the sources of asymmetry are exclusively present in $t\bar{t} + \gamma$ events where the photon originates from either an initial-state parton or one of the top quarks, events in which the photon arises from any of the charged decay products of the $t\bar{t}$ system are classified as background. To effectively discriminate between signal and background processes, a neural network (NN) is employed. The NN's output distribution is strategically leveraged to define two distinct regions: one enriched in background events and the other dedicated to signal events. 

\begin{figure}[!h!tbp]
\centering
\includegraphics[height=2.8in]{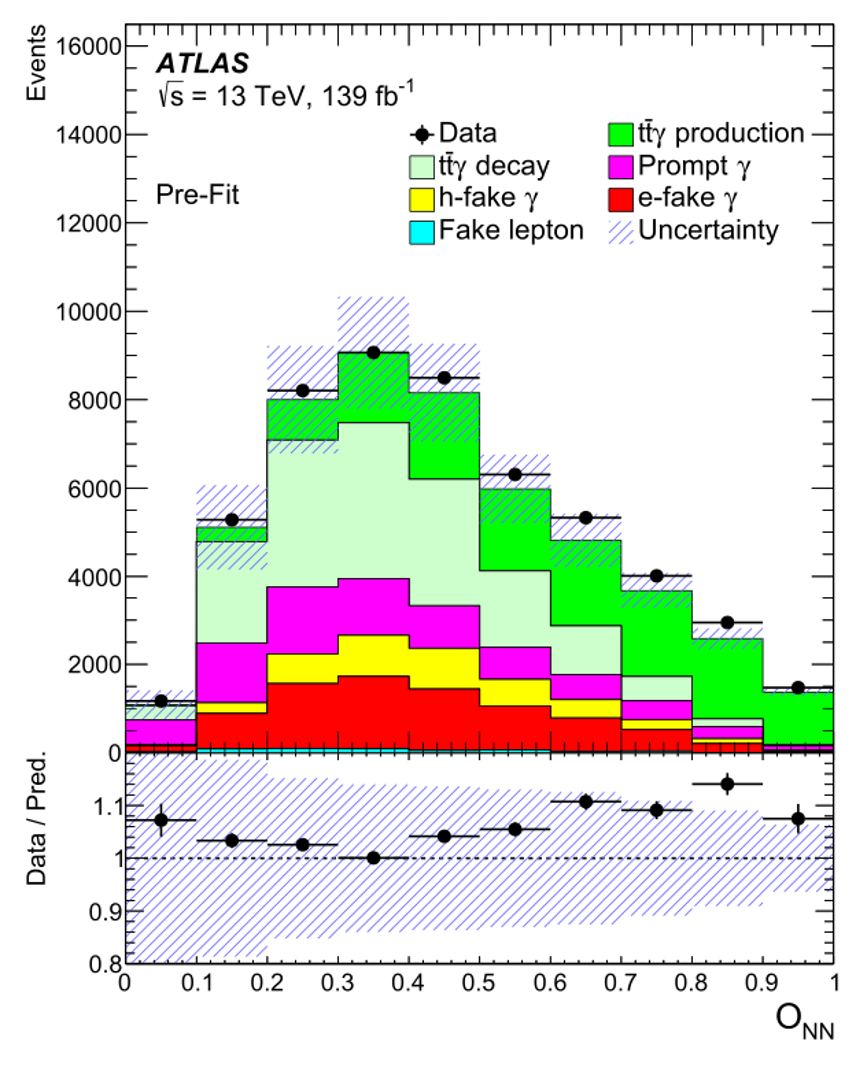}
\caption{NN output discriminant distribution before the fit. The uncertainty includes experimental and modelling systematic uncertainties. The lower part of the plot shows the ratio of the data to the prediction \cite{2023137848}.}
\label{fig:3}
\end{figure}

The value of $A_C$ is determined through a maximum-likelihood fit to the distribution of the absolute rapidity difference between the top quark and antiquark. The measurement yields $A_C = -0.003 \pm 0.029$, in good agreement with the SM expectation. The precision is constrained by statistical uncertainty (Fig. \ref{fig:3}).

\section{Measurement of the $t\bar{t}$ charge asymmetry in events with highly Lorentz-boosted top quarks in $pp$ collisions at $\sqrt{s}$ = 13 TeV}

\begin{figure}[!h!tbp]
\centering
\includegraphics[height=2.8in]{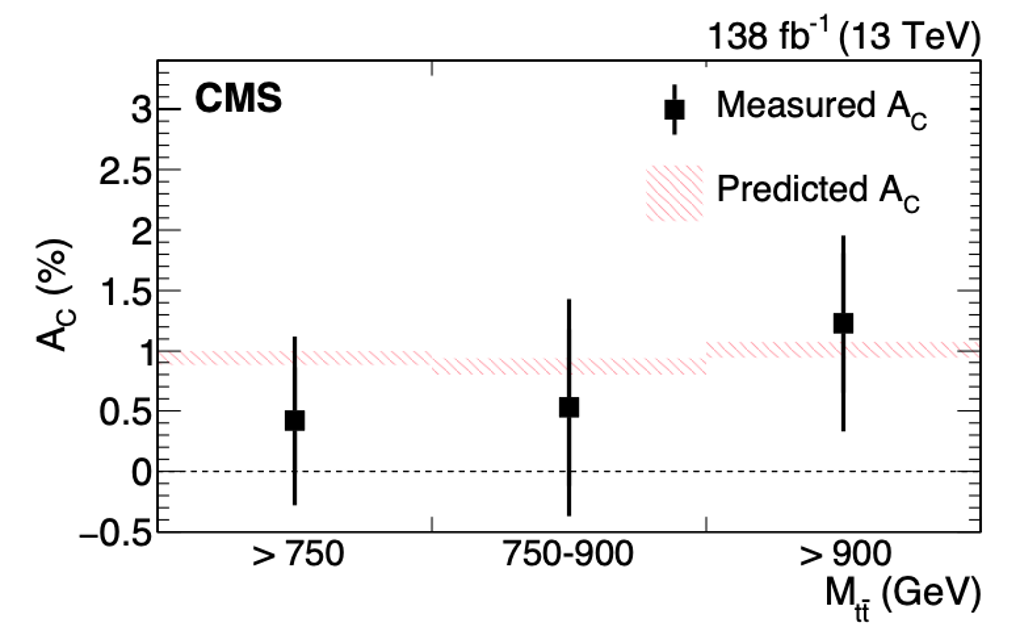}
\caption{Unfolded $A_C$ in the full phase space presented in different mass regions. The vertical bars represent the total uncertainties. The measured values are compared to the theoretical prediction, including NNLO QCD and NLO EW corrections \cite{2023137703}.}
\label{fig:4}
\end{figure}

Another approach to enhancing charge asymmetry at the LHC involves examining highly boosted $t\bar{t}$ events. Because the relative contribution of valence quarks increases at high momentum transfer, it is expected to have more $t\bar{t}$ pair initiated in $q\bar{q}$ interactions. In \cite{2023137703}, the event selection is optimized to reconstruct hadronically and leptonically decaying boosted top quarks, using state-of-the-art techniques to identify non-isolated leptons, and overlapping jets. The measurement of the top quark charge asymmetry is conducted for events with a $t\bar{t}$ invariant mass greater than 750 GeV and corrected for detector and acceptance effects using a binned maximum likelihood fit. The measured top quark charge asymmetry is $A_C = 0.42 \pm 0.64\%$ . Furthermore, the $A_C$ is also presented for two invariant mass ranges, specifically 750 -- 900 GeV and $>$ 900 GeV. The inclusive and differential measurements are in strong agreement with the standard model predictions, specifically at the next-to-next-to-leading order in quantum chromodynamic perturbation theory with next-to-leading-order electroweak corrections (Fig. \ref{fig:4}).

\section{Measurement of the polarisation of $W$ bosons produced in top-quark decays using dilepton events at $\sqrt{s}$ 13 TeV with the ATLAS experiment}

In \cite{2023137829}, a measurement of the $W$ boson polarization in top-quark decays is presented. This measurement utilizes proton-proton collision data at a center-of-mass energy of 13 TeV. The analysis focuses on $t\bar{t}$ events where both quarks decay leptonically. The key variable, the helicity angle $\theta^*$, is defined as the angle between the charged lepton and the reversed momentum direction of the b quark stemming from the top decay, both calculated in the $W$ boson rest frame.

The $W$ bosons can exhibit three types of polarization: longitudinal ($F_0$) or left-handed ($F_L$), while right-handed polarization ($F_R$) is strongly suppressed within the Standard Model. The presence of a significant $F_R$ contribution would indicate potential new physics.

\begin{figure}[!h!tbp]
\centering
\includegraphics[height=2.8in]{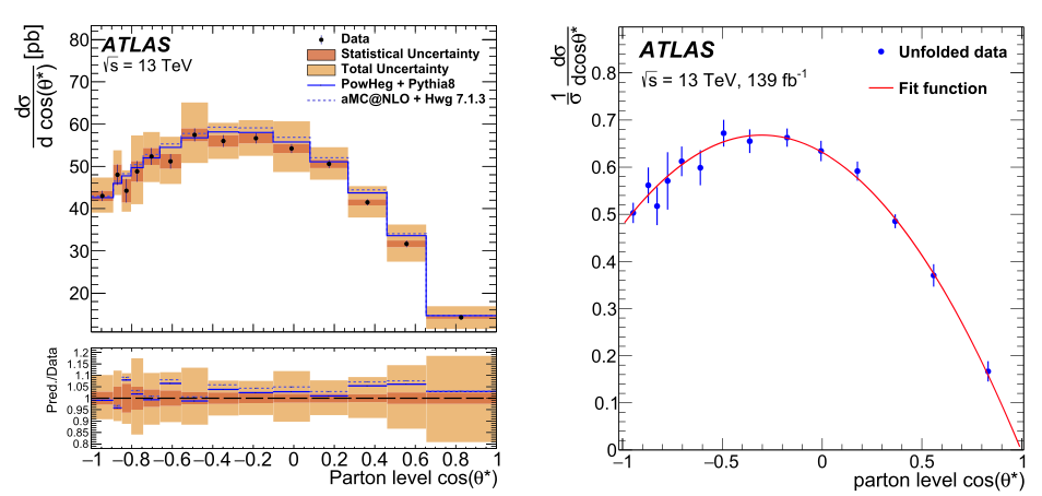}
\caption{The unfolded differential distribution (left) and the unfolded normalised distribution (right). The parton-level distribution predicted by Powheg Box interfaced with Pythia is shown. The unfolded normalised distribution in data is shown with the function of Eq. (1) overlaid, using the helicity fractions $F_0$, $F_L$ and $F_R$ determined from the fit. The total uncertainties are shown on data points \cite{2023137829}.}
\label{fig:5}
\end{figure}

The results are unfolded back to the parton level and corrected for detector resolution and acceptance. The measured polarization fractions are as follows: $F_0 = 0.684 \pm 0.005 (stat.) \pm 0.014 (syst.)$, $F_L = 0.318 \pm 0.003 (stat.) \pm 0.008 (syst.)$, and $F_R = -0.002 \pm 0.002 (stat.) \pm 0.014 (syst.)$. These measurements align with SM predictions (Fig. \ref{fig:5}).

\section{Searches for violation of Lorentz invariance in $t\bar{t}$ production using dilepton events in proton-proton collisions at $\sqrt{s} = $ 13 TeV}

\begin{figure}[!h!tbp]
\centering
\includegraphics[height=1.9in]{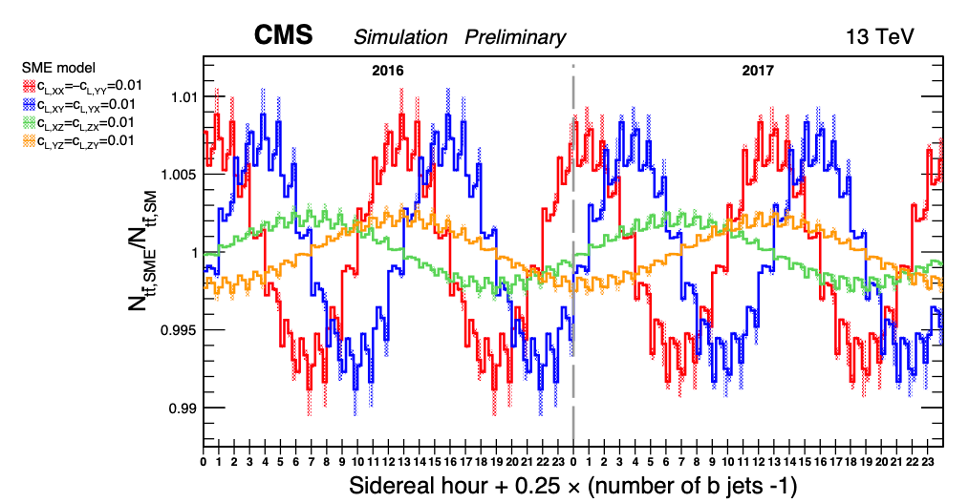}
\includegraphics[height=1.9in]{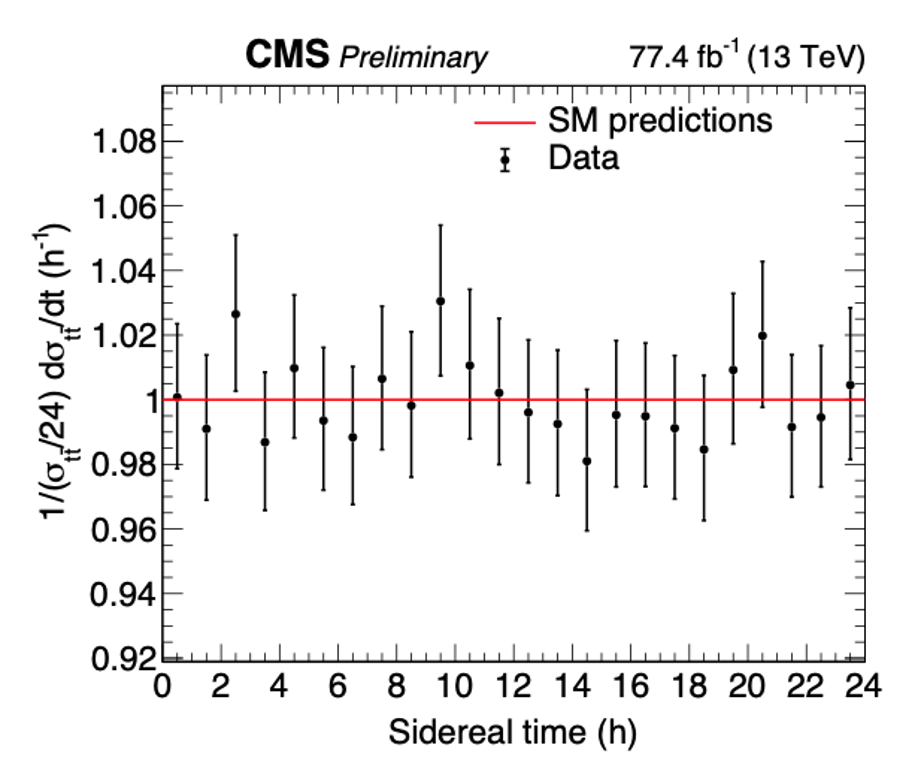}
\caption{Number of $t\bar{t}$ events reconstructed in the SME hypothesis divided by the SM hypothesis, as a function of the number of b jets and sidereal time, here for the four directions of the $t\bar{t}$ non-invariant Lorentz coefficients (left) and the $t\bar{t}$ normalized differential cross section as a function of sidereal time, using combined 2016 and 2017 data (right) \cite{WinNT}.}
\label{fig:6}
\end{figure}

Lorentz-violating Standard Model Extension (SME) is tested in the top quark sector in \cite{WinNT}, motivated by string theory and quantum loop gravitation. This has been tested in many sectors but only once with top quarks at D0 \cite{PhysRevLett.108.261603}. This is the first search for Lorentz invariance violation with top quark at the LHC using dileptonic $e\mu$ events from $t\bar{t}$ events in 77.4 fb$^{-1}$ (2016 -- 2017 data) (Fig. \ref{fig:6}). The CMS detector is moving around the earth's rotation axis during a sidereal day, and so does the beam line direction at the interaction point, or the average direction of top quarks produced in the collisions. Consequently, the time-dependent top quark couplings in the SME will result in a cross section for top quark production modulated with sidereal time.  The normalized differential cross section for $pp \rightarrow t\bar{t}$ production is measured as a function of sidereal time, in bins of hours within the sidereal day. The discriminant observable between $t\bar{t}$ and background processes is built with the distribution in the number of b jets in each sidereal time bin. Bounds on Lorentz-violating couplings are extracted, and found to be compatible with Lorentz invariance with an absolute precision of 0.1 -- 0.8\%.

\bibliography{eprint}{}
\bibliographystyle{unsrt}
 
\end{document}

%% file: econfmacros.tex
\def\beq{\begin{equation}}
\def\eeq#1{\label{#1}\end{equation}}
\def\eeqn{\end{equation}}

\def\beqa{\begin{eqnarray}}
\def\eeqa#1{\label{#1}\end{eqnarray}}
\def\eeqan{\end{eqnarray}}

\let\bar=\overbar

\def\Dslash{\not{\hbox{\kern-4pt $D$}}}
\def\dslash{\not{\hbox{\kern-2pt $\del$}}}

\def\msb{{\bar{\ssstyle M \kern -1pt S}}}

